# An information theoretic model for the linear and nonlinear dissipative structures in irradiated single-walled carbon nanotubes

*Afshan Ashraf[1,2], Sumera Javeed[2], Sumaira Zeeshan[2], Kashif Yaqub[2] and Shoaib Ahmad[3,\*]*

[1]Pakistan Institute of Engineering and Applied Sciences (PIEAS), P. O. Nilore, Islamabad 45650, Pakistan

[2]Pakistan Institute of Nuclear Science and Technology (PINSTECH), P. O. Nilore, Islamabad 45650, Pakistan

[3]National Center for Physics, QAU Campus, Islamabad 44000, Pakistan

Corresponding Author: *E-mail: sahmad.ncp@gmail.com

Abstract

Experiments with irradiated single-walled carbon nanotubes are shown to generate a set of probability distribution functions and to derive a set of information theoretic entropy-based parameters. Energetic $Cs^+$ ions initiate linear collision cascades and nonlinear thermal spikes in single-walled carbon nanotubes. The probability distribution functions are constructed from the normalized experimental yields of the sputtered atoms and clusters. The information or Shannon entropy and fractal dimension are evaluated for each of the emitted species. Along with the fractal dimension, the information is used to identify and distinguish the energy dissipation processes that generate conditions for monatomic sputtering and clusters emissions.

# Introduction

Emission of multi-atomic carbon clusters $C_x$; $x > 1$ from the heavy ion irradiated fullerenes [1], single [2] and multi-walled carbon nanotubes [3], is a rule rather than the exception. Monatomic carbon $C_1$ is either not sputtered at low ion energies or has very low yield as compared with the clusters' yields. Neither the phenomenon of cluster emission nor their consistently enhanced yields can be explained by the linear collision cascade theories [4-6]. We recently developed a thermal model to provide justifications for localized thermal spikes in irradiated nanotubes [7], Si, Ge and ZnO [8]. However, the ambiguity still persists regarding the details of the processes and sequences of ion-induced kinetic and thermal effects that produce monatomic $C_1$ on the one hand and clusters on the other. Here we analyze the results of mass spectrometry of the sputtered atoms and clusters. It is shown that $C_1$ sputtering yields have direct dependence on $Cs^+$ energy $E(Cs^+)$. This being the evidence of the existence of binary collision cascades [5,6]. The observed cluster yields are independent of the ion energy. At lower $E(Cs^+)$, the normalized sputtering yields of clusters are consistently higher than the monatomic yield [1-3]. In this letter, the irradiated SWCNTs are considered dissipative dynamical systems [9-11]. Such a dynamical system can be described by evaluating information or Shannon entropy [12] of all the dynamical processes. To identify the relative contributions of the mechanisms of cascades and thermal spikes, the normalized probabilities of emission of sputtered species $p(C_x)$; $x \geq 1$ are calculated from experimental data of SWCNTs with 2 nm diameters, irradiated with $Cs^+$ with $E(Cs^+)$ = 0.2 to 2.0 keV. From the experimentally determined probabilities, the instantaneous and cumulative entropies [12-14] of all sputtered species are obtained. We show that the fractal dimensional analysis [15-17] based on information [18] provides the information about the simultaneous existence and the relative contributions of the twin-mechanisms of binary collision cascades and thermal spikes. The fractal dimension of the four characteristic sputtered species ($C_1^-, C_2^-, C_3^-, C_4^-$) is used as a diagnostic tool that identifies cascades and thermal spikes and the associated mechanisms of energy transfer by Cs$^+$ ions to atoms of SWCNTs. We show that fractal analysis is a powerful tool to probe the ion-solid interactions.

Ion-induced sputtering is the major source of radiation damage in solids. It is a well-established field with numerous crowning technological achievements. A significant proportion of the ion energy is assumed to be dissipated in binary collision cascades. The atomic sputtering and the associated vacancy generation are defined as linear energy dissipation mechanisms. Under certain circumstances, the nonlinear component of the irradiating ion's energy is effectively operative and can generate localized thermal spikes whose experimental evidence, theoretical interpretations and MD simulations have established the conditions for various ion-target combinations and the ion-energy dependence. Comprehensive summaries of the individual, combined and the relative effects of the cascades and spikes has been provided in various reports and reviews [19-22]. The irradiation effects in carbon nanostructures have been investigated in recent years due to the potential applications of graphene and nanotubes [23,24]. In this communication, a model is developed based on the mass spectrometry of the sputtered atoms and clusters that are emitted from irradiated single-walled carbon nanotubes (SWCNTs). In this model the apparently distinct, different and diverse fields of Radiation Damage, Information theoretic-entropy and Fractal dimensional analysis have been integrated by using the appropriately defined probability distribution function $p(C_x)$ derived from the normalized emission densities of carbon atoms and clusters $C_{x\geq 1}$ from Cs$^+$-irradiated SWCNTs.

## Materials and methods

The Source of Negative Ions by Cesium Sputtering-(SNICS) [24] and a momentum analyzer were used for experiments to study the sputtering of SWCNTs. The parameters of the Cs$^+$ beam can be accurately controlled. The energy of the Cs$^+$ ions E(Cs$^+$) was varied between 0.2 and 2.0 keV. The Cu bullets containing SWNTs of 2 mm diameter and 3-13 micrometer length were used as targets for NEC's SNICS II negative ion source mounted on the 2MV Pelletron at Center for Advanced Studies in Physics (CASP) at Government College University (GCU), Lahore. The total beam energy was maintained by selecting the combination of the extraction and acceleration voltages at 20 kV. This ensured that the sputtered ions are extracted with the same energy even though these were subjected to varying Cs$^+$ energies E(Cs$^+$). The high electron affinities of the sputtered species facilitate their detection. A 30° bending magnet analyzed the anions. Detection of negative anions has its advantages over that of the positively charged cations. Hot plasmas are required for the production of positive charges to remove at least one electron from the respective species. Larger clusters are likely to fragment in collisions with hot electrons and other charged ions [25]. SNICS operates at low temperature (~100°C) at which neutral Cs$^o$ ionizes. It is a preferable, low temperature collision chamber to study the sputtered neutral species by conversion into anions with the attachment of electrons after ejection from the irradiated surface.

## The probability distribution function $p(C_x)$

Mass spectra of anions is obtained for an extended range of Cs$^+$ energies-$E(Cs^+)$ from low to moderate energies (~ 0.2 to 2.0 keV) in 0.1 keV steps. The normalized emission probability of C$_x$ as a function of $E(Cs^+)$ is defines as

$$p(C_x) \equiv p_{C_x}(E(Cs^+)) \qquad (1).$$

In this communication $p(C_x)$ with the implied $E(Cs^+)$ dependence will be used as shown in equation (1). The advantages of using this probability distribution function in the context of radiation damage theory, information theoretic entropy and fractal dimensions is to understand and explain four experimental observations from the mass spectra of the sputtered atoms and clusters $C_{x\geq 1}$; (i) the absence of monatomic $C_1$ at very low $E(Cs^+)$ and it's consistent, low relative densities at higher $Cs^+$ energies, (ii) the higher, relative densities of $C_2$, $C_3$ and $C_4$ clusters at all irradiation energies, (iii) the energy dependence of the probability $p(C_1)$ of $C_1$ on $E(Cs^+)$ and (iv) the observation that probabilities $p(C_x)$ of $C_2$, $C_3$ and $C_4$ do not have explicit dependence on $E(Cs^+)$. These experimental observations are vividly demonstrated by the data in Figure 1(a) that has the mass spectrum dominated by clusters $C_{x>1}$. $C_1$ has very low yields. Figure 1(b) clearly demonstrate this fact with low $p(C_1)$ as compared with $p(C_2)$ and the probabilities of the higher clusters. Figure 1(b) shows the probability distribution function as the basic data content of the radiation damage-related description of the irradiated SWCNTs.

Figure 1(a) shows the mass spectrum of carbon anions sputtered at $E(Cs^+) = 0.8$ keV. Figure 1(b) has the cumulative data of emission probabilities $p(C_x)$ of the six sputtered species of the entire energy range of $Cs^+$. The data presented in figure 1(b) is for anions whose structural stability is well established [26]. However, we further tested the stability of individual anions ($C_1^-$ to $C_6^-$) by accelerating each individual carbon anion to the High Voltage Terminal of 2 MV Pelletron [27]. Anions' structural stability after emission from SWCNTs is essential for accurate analysis and the conclusions reached thereafter. The normalized probability profiles of the five negatively charged sputtered clusters $C_x^-$, $x > 1$, and the atomic $C_1^-$ shown in Fig. 1(b), are obtained from the areas under each species from the set of nineteen mass spectra. $C_2^-$, $C_3^-$ and $C_4^-$ have the largest contributions at all ion energies. At $E(Cs^+) = 0.2$ and 0.3 keV, these three are the only sputtered species. The first appearance of $C_1^-$ is at $E(Cs^+) = 0.4$; thereafter it makes gradual, consistently increasing contributions as a function of $E(Cs^+)$.

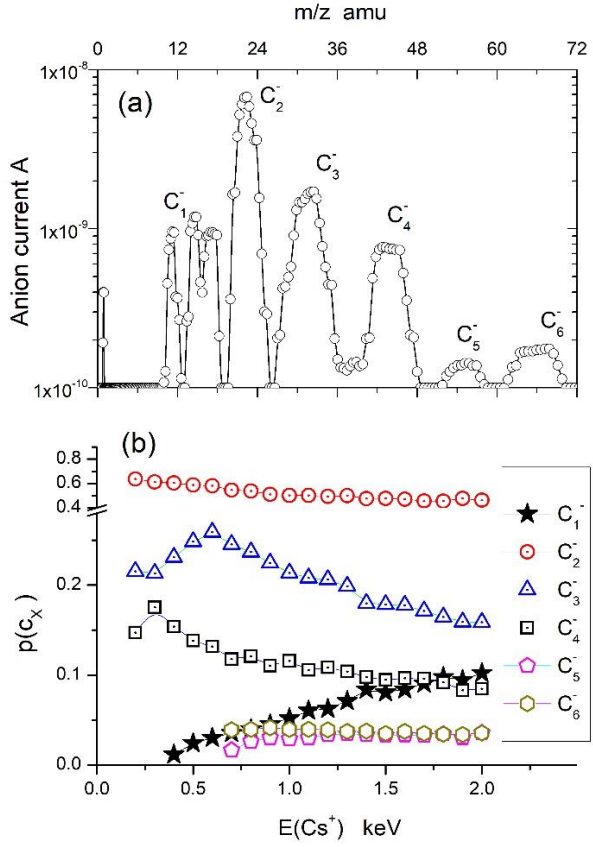

**Figure 1.** (a) Mass spectrum of $C_1^-$, $C_2^-$, $C_3^-$, $C_4^-$, $C_5^-$ and $C_6^-$ sputtered from SWCNTs at E(Cs$^+$) = 0.8 keV is shown. (b) Probabilities $p(C_x)$ of emission the six anions, from nineteen spectra are plotted with Cs$^+$ energy steps $\epsilon = $ 0.1 keV for the energy range E(Cs$^+$) = 0.2 and 2.0 keV. **Inset**: shows the symbols used for each species. The vertical axis in 1(b) shows a break to expands the region where $C_1^-$, $C_5^-$ and $C_6^-$ have low and overlapping probabilities.

## Information theoretic entropy and fractal dimension

We evaluate the information theoretic entropy $I_x$ from the experimentally determined probability distribution $p(C_x)$ as a function of $E(Cs^+)$ as

$$I_x \equiv H(p(C_x)) = - \sum_\epsilon p(C_x) \ln p(C_x) \qquad (2).$$

The summation is over all energy steps $\epsilon$ of the probability distribution for each sputtered species, from the minimum to the maximum $E(Cs^+)$. The information theoretic analysis is used to identify the processes where the external energetic ions induce radiation damage in irradiated nanotubes. Entropy $I_x$ for each sputtered species provides a measure of information about the underlying mechanism that is responsible for the emission of atoms and clusters. The output from the irradiated nanotube is treated as the information given out from the SWCNT that received a well-defined ionic energy input.

In Figure 2(a) and (b), the radiation damage aspect of the model is demonstrated. The probabilities $p(C_x)$ for the pristine and damaged SWCNTs are derived from the data of anion currents obtained from the two sets of 19 mass spectra each, with $\epsilon = 0.1\ keV$. Information theoretic entropy $I_x$ is the sum of $-p(C_x)lnp(C_x)$ versus $E(Cs^+)$ shown in 2(c) and 2(d). It is obtained from the probability data for the pristine and the damaged sets of SWCNTs.

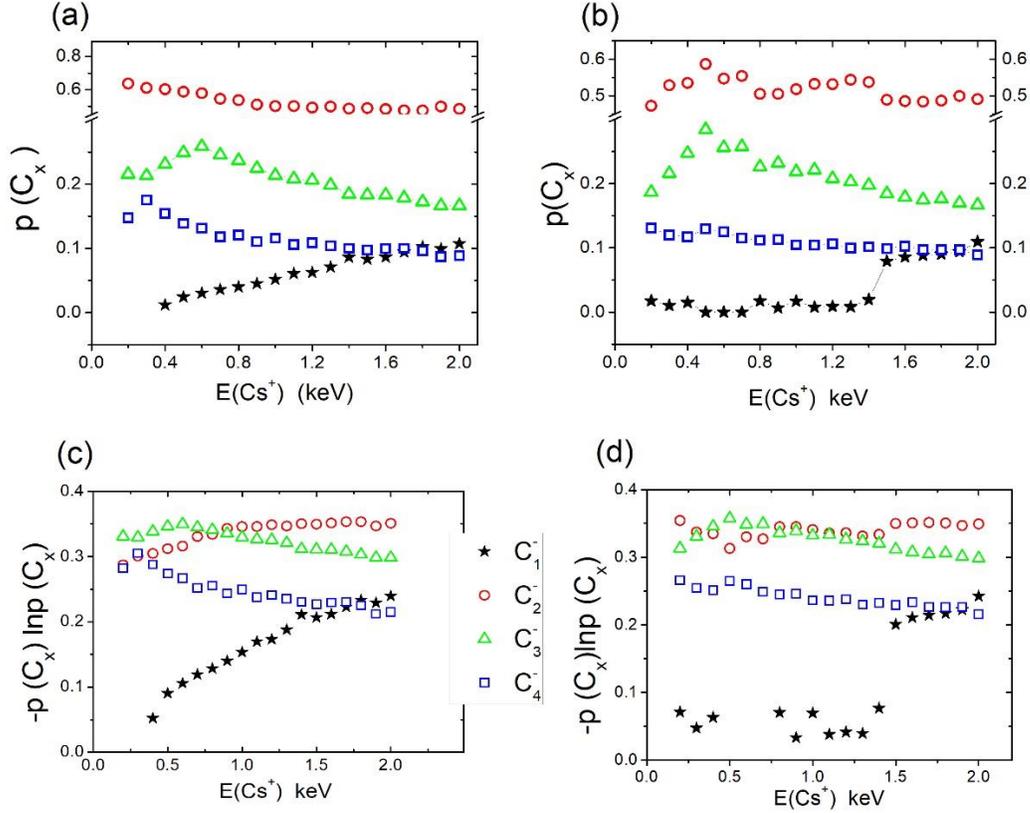

**Figure 2.** (a) The probability distributions $p(C_x)$, for the pristine sample of SWCNTs, of the 4 emitted species $C_1$, $C_2$, $C_3$ and $C_4$ are shown from the normalized anion currents at each step of $E(Cs^+)$. (b) For the highly irradiated and damaged set of SWCNTs, the probability distributions are plotted against $E(Cs^+)$. (c) The entropy $-p(C_x)lnp(C_x)$ for each of the species, at every energy step $E(Cs^+)$ is plotted for the pristine sample with $p(C_x)$ obtained from 2(a). Figure 2(d) has the same for the highly irradiated SWCNTs shown in (b).

The information $I_x$ alone cannot provide conclusive evidence about the detailed character of the energy dissipation mechanisms. We show that fractal dimension derived from information $I_x$ emerges as an additional, analytical tool to unambiguously characterize the mechanisms responsible for atomic and cluster emissions. Fractal dimension can be calculated from the plots of $-p(C_x)lnp(C_x)$ against $E(Cs^+)$ as

$$d_f(C_x) = -\sum_\epsilon p(C_x) \ln p(C_x)/\ln(1/\epsilon) \quad (3).$$

In figure 3(a) the fractal dimensions $d_f(C_x)$ are evaluated for the sputtered species from $C_1$ to $C_4$. The two parameters $I(C_x)$ and $d_f(C_x)$ are employed in this communication to identify, distinguish and quantify the externally induced dynamic processes in irradiated carbon nanotubes. The above mentioned four (4) experimental observations from the mass spectrometric data and its information theoretic description based on $I(C_x)$ and $d_f(C_x)$ configures the two dynamic processes; the linear collision cascades and the nonlinear thermal spikes. Two different energy scales are associated with the two processes that will be discussed later.

Another important parameter, relative entropy $D((p(C_1) \| p(C_2))$ is defined using the probability distributions of $C_1$ and $C_2$ to estimate the distance between the two physical processes that generate them. Relative probability is a measure of the distance between the probability distributions $p(C_1)$ and $p(C_2)$. In physical terms, it implies that the two sets of information may have different origins or spatial properties. It is defined in ref. [14] as

$$D(p(C_1) \| p(C_2)) = p(C_1) \ln(p(C_1)/p(C_2)) \quad (4).$$

In figure 3(b), the relative probability of $C_1$ versus $C_2$ reduces by 50% for the damaged SWCNTs as opposed to the value for the pristine nanotubes. It implies that the distance between the two probability distributions $p(C_1)$ and $p(C_2)$ reduces in the damaged SWCNTs.

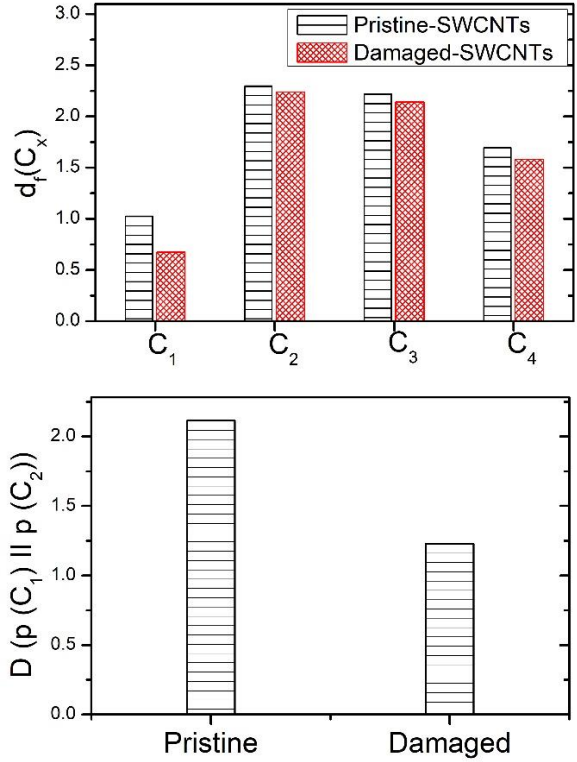

**Figure 3.** The fractal dimensions of the four emitted species $C_1$, $C_2$, $C_3$ and $C_4$ are plotted for the two sets in (a) of the pristine and damaged SWCNTs. (b) Relative entropy $D(p(C_1) \parallel p(C_2))$ as a measure of the distance between the two probability distributions of $C_1$ and $C_2$ is calculated and shown for the pristine and damaged SWCNTs in (b).

The qualitative picture of $C_1$'s dependence on $Cs^+$ energy and the non-dependence of $C_2$, $C_3$ and $C_4$, is obvious from the $p(C_x)$ versus $E(Cs^+)$ plots in 2(a) and 2(b). Similar message is conveyed by the entropy plots in 2(c) and 2(d). The quantitative analysis is presented in figure 3(a) and 3(b). For $C_1$ the fractal dimension $d_f(C_1) \sim 1$ for the pristine and $< 1$ for the damaged SWCNTs. Fractal dimension is derived in equation (3) from the information $I_1 \equiv I(C_1) = - \sum_\epsilon p(C_{1)}) \ln p(C_1)$ that emerged from the energy dissipation mechanisms with $E(Cs^+)$ dependence. The pattern of $I_1$ as a function of $E(Cs^+)$ and $d_f(C_1) \lesssim 1$ is indicative of linear, non-space-filling fractal-the collision cascades. A recent experimental investigation [28] that utilized and compared their data with the Monte Carlo simulations of SRIM [29] has confirmed the fractal dimension of monovalent vacancies $d_f(single-vacancies) \sim 1$. Our experimentally determined fractal dimension for $C_1$ is shown in figure 3(a) as $d_f(C_1) \sim 1$.

For $C_2$ and higher clusters $d_f(C_{x \geq 2}) \sim 2$ in figure3(a). Fractal dimension $\sim 2$ implies a space-filling fractal which constitutes a localized region that emits diatomic and larger clusters. It can happen only if the local temperature is high $\sim T_s$ and the energies of formation of multi-vacancies is less than the energy required for a monovalent vacancy $E_{xv}[x \geq 2] < E_{1v}$. This is the essential requirement for $p(C_2) > p(C_1)$. Only a space-filling, multifractal, localized thermal spike can describe $d_f(C_2) \sim d_f(C_3) \sim 2$. The conditions that generate such a spike in irradiated SWCNTs have been discussed in ref. [7.]. The probability of emission of a cluster $C_x$ with energy of formation $E_{xv}$ for an x-valent vacancy formation at temperature $T_s$ is estimated as

$$p(C_x) \propto (exp(E_{xv}/T_s) + 1)^{-1} \qquad (5).$$

In this equation $p(C_x)$ is not dependent on $Cs^+$ energy but on $T_s$. The $p(C_x)$ vs $E(Cs^+)$ spectra in figure 2(a) and 2(b) confirms this conclusion. For the highly irradiated and damaged SWCNTs $d_f(C_1) < 1$. This may be the result of disruption of the C-C bonding due to creation of a significant number of 2-, 3- and 4-atom vacancies on the hexagonal network. On the other hand, the spike-related multi-atomic cluster emissions are equally proficient in damaged SWCNTs as can be seen in figure 2(b) and 2(d).

We must emphasize that information theoretic entropy is a measure of ignorance, therefore, the relative entropy provides information for the outputs of the two mechanisms of linear cascades and nonlinear spikes that are not be directly related. Their representative probability distributions are distant from each other. This, relative distance is shown to be farther in pristine SWCNTs as opposed to that in the damaged ones. The curvature and large aspect ratio of length to diameter in nanotubes are responsible for the fundamental difference between the nature of defect formation in SWCNTs and graphite. The effects related with the radial dissipation and confinement of the kinetically received energy are more pronounced over the axial effects. One of the results is the relatively higher formation energies of single vacancy ~ 7 eV as compared with~ 3-4 eV for a divacancy in SWCNTs [28, 29] . As a result, nanotubes with 2-3 nm diameter resist formation of dangling bonds associated with single vacancies. This fact restricts the emission of an atom struck with less than the threshold energies. The kinetic energy received by such an atom is shared with its neighbors. The finite size of SWCNTs modifies the mechanism of conversion of electronic excitations into atomic kinetic energy. The lifetimes of the excitation are generally longer in nanotubes than those in graphite. Radially confined, localized excitations populate antibonding orbital that may contribute to defect formation and thermal effects [24]. We are proposing such a mechanism in figure 4.

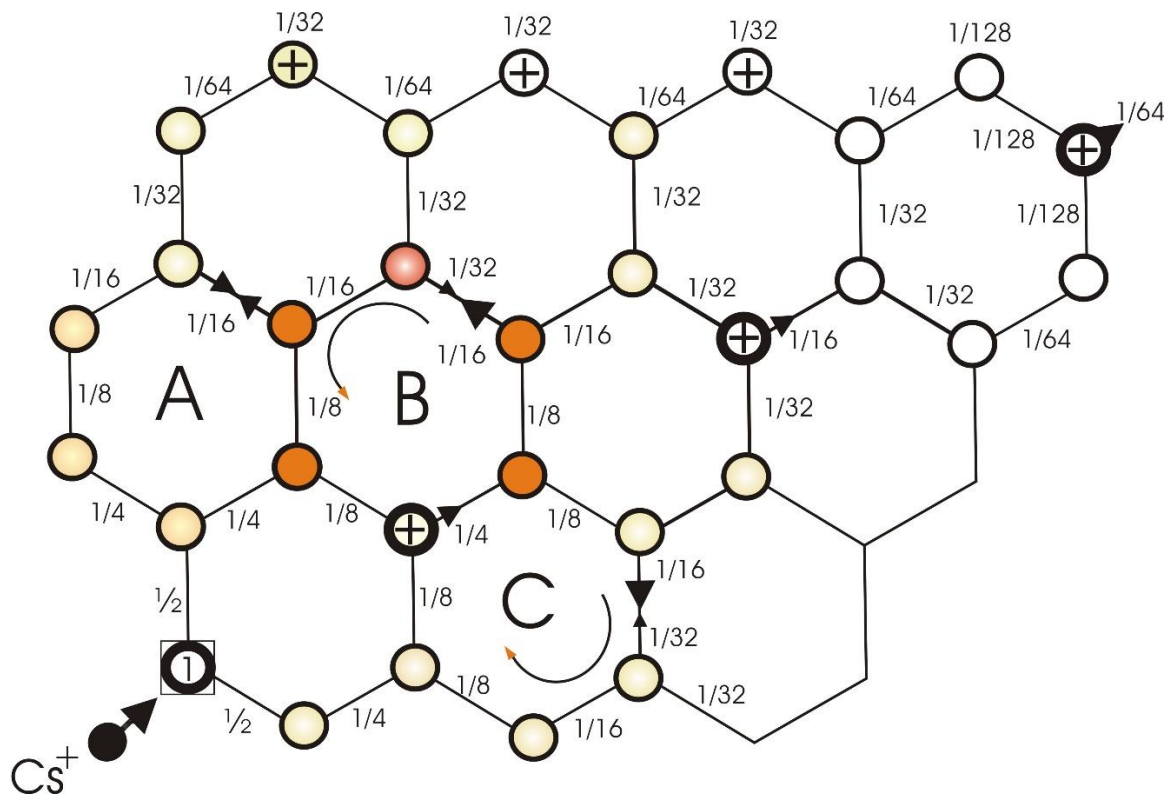

**Figure 4.** The origin of thermal spikes in SWCNTs. Hexagonal networks are shown with the first struck atom labelled as 1 and boxed. The cycle of binary energy sharing with neighboring atoms is shown with received energies along the bonds. Energy recycling is shown in the rectangles A, B and C. Symbols and arrows are described below.

Figure 4 highlights the essential features of energy sharing by the sp²-bonded atoms of the hexagons. The fundamental difference between the energy dissipation mechanisms in bulk solids and SWCNTs is due to the covalently bonded carbon atoms in hexagons on a mono-shelled surface. The figure shows a typical hexagonal network. The first atom receives energy $E_1 \equiv 1$. If $E_1 \gg E_{dis}$—the energy required to break all bonds and to leave its lattice site, then a typical binary collision cascade is generated with the energetic recoiling atom. The collisions proceed until recoiling energy $<E_{dis}$. This is the well-known cascade theory [4-6]. It is also the basis of Monte Carlo simulations SRIM [31]. Each collision generates a Frankel pair of interstitial atom and a vacancy. However, the concept of interstitials is valid only in bulk solids, these become the sputtered atoms in the case of SWCNTs.

If $E_1 \lesssim E_{dis}$ then there are three different processes of energy sharing among atoms of the hexagonal network. The first is indicated in figure 4 by the plus (+) sign on the three atoms shown with arrows. The three labelled atoms receive energy $\left(\frac{1}{2}\right)^n$ with $n \geq 2$. This is the energy spreading outwards and away from the point of the initial impact. The second mechanism is shown by the opposing arrows, like a bow-tie, in the hexagon labelled A. In this case, the atoms on the opposing edges of the hexagon receive equal energies, in opposite directions. These collisions increase the vibrational frequencies by localizing energy in collisions with the surrounding atoms. These collisions localize the deposited energy rather than spread it outwards. The third mechanism is shown by two unequal, opposing arrows on the interatomic bonds in hexagons B and C. These are due to collisions between atoms which deliver, or share, different amounts of energies, like 1/16 and 1/32 in the two cases. The excess energy is directed towards the atom with lesser energy. It is further recycled in hexagons B and C as shown by curved arrows. The latter two processes increase the vibrational energies and consequently the local temperature. It eventually generates local disorder with average atomic energies $\sim \frac{1}{5}$ to $\frac{1}{3} eV$. The spike temperatures $T_s \sim 3500K$ can be generated and have been calculated from the sputtered cluster probabilities [7]. At these temperature clusters are emitted from the localized subliming regions. Depending upon the initial energy, any of the atoms shown with (+) sign or arrows, can be sputtered. Whenever that happens, atomic sputtering is proposed to occurs and identified as $C_1$ peak in the mass spectrum in Figure 1(a). The experimentally observed $C_1$ yields have the direct dependence on ion energy.

## Conclusions

In conclusion, an information theoretic model for the fragmenting, irradiated SWCNTs has been described that has the calibrated $Cs^+$ energy as the input signal. The energy is consumed and dissipated in linear and nonlinear processes. The output signal appears in the form of atoms and clusters sputtered from the surface of the irradiated nanotube. Probability distribution functions are constructed, for each emitted species, from their normalized current densities. Information is compiled for all emitted constituents as Shannon and the relative entropies. The information so obtained, is employed to calculate fractal dimension of the sputtered species. We have shown that together, these information theoretic parameters identify, distinguish and characterize the existence and the relative operational efficiency of the linear cascades and nonlinear thermal spikes on the surface of the irradiated SWCNTs. The information theoretic entropic model developed for the irradiated carbon nanotubes can be extended to the energy dissipating environments where both the linear and nonlinear processes are operative.